\documentclass[12pt]{iopart}
\expandafter\let\csname equation*\endcsname\relax
\expandafter\let\csname endequation*\endcsname\relax

\usepackage{appendix}
\usepackage{iopams}
\usepackage{xcolor}
\usepackage[pdftex]{graphicx}
\usepackage[]{algorithm2e}
\usepackage{optidef}
\usepackage{amsthm, amsmath, amssymb}
\usepackage{comment}
\usepackage{epstopdf}

\begin{document}

\title[Steering witnesses for unknown Gaussian quantum states]{Steering witnesses
for unknown Gaussian quantum states}

\author{Tatiana Mihaescu (1,2), Hermann Kampermann (3),  Aurelian Isar (1,2),
Dagmar
Bru{\ss} (3)}

\address{(1) Department of Theoretical Physics, National Institute of Physics and
Nuclear Engineering, RO-077125 Bucharest-Magurele, Romania

(2) Faculty of Physics, University of Bucharest, RO-077125 Bucharest-Magurele,
Romania

(3) Heinrich-Heine-Universit\"at D\"usseldorf, Institut f\"ur Theoretische Physik
III, D-40225 D\"usseldorf, Germany}
\ead{mihaescu.tatiana@theory.nipne.ro}
\vspace{10pt}

\begin{abstract}
We define and fully characterize the witnesses based on second moments
detecting
steering in Gaussian states by means of Gaussian measurements. All such tests,
which arise from linear combination of variances or second moments of canonical
operators, are easily implemented in experiments. We propose also a set of
linear
constraints fully  characterizing steering witnesses when the steered party has
one bosonic mode, while in the general case the constraints restrict the set of
tests detecting steering. Given an unknown
quantum state we implement a semidefinite program providing the appropriate
steering
test with respect to the number of random measurements performed. Thus, it is a
"repeat-until-success" method allowing for steering detection with less
measurements than in full tomography. We study the
efficiency of steering detection for two-mode squeezed vacuum states, for
two-mode general
unknown states, and for three-mode continuous variable GHZ states. In addition,
we discuss the robustness of this method
to statistical errors.
\end{abstract}

\section{Introduction}

\indent The study of quantum correlations is a cornerstone of quantum information
theory enriching the foundational understanding of quantum theory and allowing
applications which outperform any classical approach in certain tasks such as
computation \cite{brus}, secure communication \cite{gras} and metrology
\cite{metr}.
Schr\"odinger \cite{schr1, schr2} discussed the enchanting phenomenon where one
party, Alice, is able to "steer" the state of a distant party, Bob, by means of
entanglement they share. The implied "action at a distance" was the core argument
in
the Einstein, Podolsky and Rosen (EPR) paper \cite{epr} against the completeness
of
quantum theory, but only lately quantum steering was conceptualized as a
particular
type of nonlocality \cite{wis}.\\
\indent In a quantum steering scenario two distant parties (Alice and Bob) share
a
common quantum state, where one of the parties, say Alice, is able to convince
Bob
that the state they share is entangled. She does so by performing local
measurements
and using classical communication, whereas Bob verifies whether the joint
probability distribution can be explained by a local hidden state (LHS) model
\cite{wis}, in which case it rules out the possibility that Alice is steering
Bob's
state by her choice of local measurement settings. This definition presents
quantum
steering as an intermediate form of correlation between entanglement and Bell
nonlocality. It has been intensively studied leading to some experimental
applications such as subchannel discrimination \cite{piani} and one-sided
device-independent cryptography \cite{branc}.\\
\indent In the context of continuous variable states \cite{ser} quantum steering
is
extensively investigated starting with Gaussian states and using Gaussian
measurements \cite{wis,kog}, that have a distinct role in the
infinite-dimensional
Hilbert space, being also readily available in experiments \cite{oliv,ferr,
weed}.
In this particular case, the necessary and sufficient criterion for steerability
is
given in terms of the covariance matrix of the state \cite{kog, ji, frig}, which
comprises the variances of the canonical operators. Therefore, the detection of
steering in a general unknown Gaussian state requires the full knowledge of the
covariance matrix, which might be excessive and resource-consuming. \\
\indent In this paper first we prove a criterion for Gaussian steerability that
is
equivalent to known criteria in the literature, but it unfolds very nice
properties of the
covariance matrix resembling the known covariance matrix criterion of
entanglement
(see Section 3, Theorem 2). A necessary condition proof of this theorem is
presented in Ref. \cite{ji}, using local uncertainty relations (LUR). Based on
this result we define
the set of linear tests  detecting Gaussian steering, or witnesses, which arise
as
linear combinations of the second moments of the canonical observables. Such
tests
are commonly used for the detection of entanglement, known as entanglement
witnesses
\cite{hyll, janet, tamih}. An entanglement witness based on second moments is a
real
symmetric matrix $Z\geq0$ such that ${\rm Tr}[Z\gamma_s]\geq 1$ holds for all
separable covariance matrices $\gamma_s$, while ${\rm Tr}[Z\gamma]<1$ for some
entangled covariance matrix $\gamma$ \cite{hyll}. An analogous definition holds
also
for steering witnesses \cite{ma} due to the  fact that the set of non-steerable
covariance matrices is convex and closed. We introduce constraints fully
characterizing the set of
Gaussian steering witnesses, which are shown to be stronger than analogous
constraints on
entanglement witnesses. \\
\indent In addition, we propose a set of linear constraints that are stronger
than the constraints fully characterising the steering witnesses, however in the
particular case, when the steered party
has just one bosonic mode, these new constraints fully characterize the set of
steering
witnesses. This allows us to write a semidefinite program finding the
optimal steering test for a given state.  We analyze the efficiency of steering
detection in unknown covariance matrices with respect to the number of random
measurements required for this task. For entanglement detection based on
covariance
matrices an analogous method was developed in Ref. \cite{tamih}, and therefore,
the
results in this article provide us with a framework of comparison between the
detection of Gaussian steering and entanglement.\\
\indent  The paper is organized as follows: Section 2 gives an introduction to
the
notions of Gaussian states, Gaussian measurements and symplectic transformations.
In Section 3 quantum Gaussian steering is defined and the covariance matrix
criterion for Gaussian steering is proven. In Section 4 the steering witnesses
based
on second moments are introduced and fully characterized. In Section 5 we
construct steering witnesses from random measurements acquired by homodyne
detection. Section 6 presents the results of steering detection in two-mode
squeezed vacuum
states and in two-mode general unknown states. Also, we illustrate the example of
steering detection in three-mode continuous variable GHZ states, where the
steered party has two modes. In Section 7 the statistical analysis of
our method is provided. The summary and conclusions are expanded in Section 8.

\section{Gaussian states}

A continuous variable (CV) system of $N$ bosonic modes is described by the
canonical
operators of position and momentum $\hat x_k$ and $\hat p_k$ in the Hilbert space
$
\mathcal H=\bigotimes_{k=1}^N \mathcal H_k$, where $H_k$ is the
infinite-dimentional
Hilbert space of mode $k$ \cite{ser, oliv,ferr}. Defining the vector of canonical
operators $\mathbf{\hat R}^{\rm T}\equiv(\hat R_1,...,\hat R_{2N})=(\hat x_1,\hat
p_1,...,\hat x_N,\hat p_N)$ one obtains the commutation relation written as (we
assume $\hbar=1$):
\begin{equation}\label{cmr}
  [\hat R_i,\hat R_j]={\rm i}\Omega_{ij} {\rm \hat I},\quad i,j=1,...,2N,
\end{equation}
where $ {\rm \hat I}$ is the identity matrix, and $\Omega_{ij}$ are the elements
of
the symplectic matrix
\begin{equation}\label{sym}
\Omega_N=\bigoplus_{1}^N \left({\begin{array}{*{50}c} 0 & 1 \\ -1 & 0 \\
\end{array}}\right).
\end{equation}
Gaussian states are fully described by the first and second order statistical
moments, namely the displacement vector $\mathbf{d}={\rm Tr}[\mathbf{\hat R}
\rho]$, and the
covariance matrix (CM) $\gamma$ with its elements defined as \cite{weed, sim}:
\begin{equation}\label{ptr}
\gamma_{ij}=\langle\{\hat R_i-\langle\hat R_i\rangle, \hat R_j-\langle\hat
R_j\rangle\}_+\rangle_\rho,
\end{equation}
where $\{ ,\}_+$ represents the anticommutator. The CM of a physical quantum
state
has to fulfill the Robertson-Schr\"odinger uncertainty relation, which we will
often
refer to as \textit{covariance matrix criterion} (CMC):
\begin{equation}\label{unc}
\gamma+{\rm i}\Omega_N \geq 0.
\end{equation}
In the following we will mostly consider a bipartite Gaussian quantum state
$\rho_{AB}$ of $N$ modes with the CM of the following block structure:
\begin{equation}\label{bl}
\gamma_{AB}=\begin{pmatrix}
\gamma_A & \gamma_{12}\\
\gamma_{12}^{\rm T} & \gamma_B
\end{pmatrix},
\end{equation}
where $\gamma_A$ and $\gamma_B$ are the CMs of the subsystems of Alice and Bob
with
$N_A$ and $N_B$ modes, respectively, $\gamma_{12}$ is the correlation matrix
between
the two parties, and $N=N_A+N_B$. It can be readily shown that the uncertainty
relation in Eq. (\ref{unc}) directly implies $\gamma_{AB}>0$, $\gamma_A>0$ and
$\gamma_B>0$.

A much more general approach in describing any $N-$mode CV state is based on the
completeness of the set of displacement operators in phase space defined as:
\begin{equation}\label{disp}
  \hat D(\mathbf{r})={\rm e}^{{\rm i}~\mathbf{r}^{\rm T}\Omega_N\mathbf{\hat R}},
\end{equation}
where $\mathbf{r}^{\rm T}=(x_1, p_1,..., x_N, p_N)$ is a real vector of phase
space
variables. The connection between Hilbert space and phase space descriptions is
given by
the Fourier-Weyl relation for a given density operator $\hat \rho$ in Hilbert
space
\cite{ser}:
\begin{equation}\label{weyl}
  \hat \rho=\frac{1}{(2 \pi)^N}\int_{\mathbb{R}^{2N}}d^{2N}\mathbf{r} ~{\rm
  Tr}[\hat D^\dag
  (\mathbf{r}) \hat \rho] \hat D(\mathbf{r}),
\end{equation}
from where the characteristic function is readily defined as
$\chi_\rho(\mathbf{r})={\rm
Tr}[\hat D^\dag(\mathbf{r}) \hat \rho]$. If $\hat \rho$ represents a Gaussian
state then its
characteristic function takes a particular form \cite{ser}:
\begin{equation}\label{gchf}
  \chi_\rho(\mathbf{r})={\rm e}^{-\frac{1}{4} \mathbf{r}^{\rm T} \Omega_N^{\rm T}
  \gamma_{AB} \Omega_N
  \mathbf{r}},
\end{equation}
where the first moments are considered to be zero and $\gamma_{AB}$ is the CM
associated with the Gaussian density operator $\hat \rho$. A Gaussian measurement
\cite{fur} applied on Alice's subsystem can be described by a positive Gaussian
operator $\hat A$ with a CM given by $T^A$ satisfying the CMC: $T^A+{\rm i}
\Omega_{N_A}\geq 0$. The calculation of ${\rm Tr}_A[(\hat A \otimes \mathcal{\hat
I}_B) \hat \rho]$, where $\mathcal{\hat I}_B$ is the identity operator defined on
Bob's space, transforms into a multivariate Gaussian integral on characteristic
function level. The remaining modes of Bob's conditioned state $\tilde{\rho}_B^A$
after measurement form a Gaussian state with CM given by
$\gamma_B^A=\gamma_B-\gamma_{12}^{\rm T}(\gamma_A+T^A)^{-1}\gamma_{12}$, which
represents the \textit{Schur complement} of the matrix $\gamma_{AB}+T^A\oplus
0_B$ with
respect to the submatrix $\gamma_A+T^A$ \cite{ser}. The following Lemma of the
Schur
complement will be useful in further discussions on Gaussian steering.\\

\textit{\textbf{Lemma 1} \cite{zhan}. Consider a Hermitian matrix
\begin{equation}\label{hm}
  H=\begin{pmatrix}
               A & X \\
                X^\dag & B
              \end{pmatrix}.
\end{equation}
Then $H>0$ if and only if $A>0$ and $H/A=B-X^\dag A^{-1}X>0$, where $H/A$ is the
Schur complement of block $A$ of the matrix $H$.
}\\

Given that $\gamma_{AB}$ is a CM satisfying the uncertainty relation in Eq.
(\ref{unc})
we have $\gamma_{AB}>0$ and $\gamma_A>0$, and hence the Schur complement of
$\gamma_{AB}$ with respect to $\gamma_A$ is also a positive matrix
$\gamma_{AB}/\gamma_A:=\gamma_B-\gamma_{12}^{\rm T}\gamma_A^{-1} \gamma_{12}>0$.
Let
us consider the matrix $\gamma_{AB}-0_A \oplus \sigma_B$, with $\sigma_B\geq 0$,
and its
Schur complement with respect to $\gamma_A$. Applying the positivity conditions
from Lemma 1 leads to the  Schur complement \cite{ser,
lami}
\begin{equation}\label{scb}
  \gamma_{AB}/\gamma_A=\sup\{\sigma_B : \gamma_{AB}\geq 0_A\oplus \sigma_B\},
 \end{equation}
which means that the matrix set on the right-hand side has a supremum (i.e. a
minimum upper bound) with respect to the L\"owner partial order ($X\geq Y$ if and
only if $X-Y$ is positive semidefinite), and that this supremum is given by the
Schur complement on the left-hand side.

\subsection{Symplectic transformations}
The equivalent of unitary operators acting on the quantum state space are the
symplectic transformations $S$ on CMs, which are symmetric real matrices acting
by
congruence on CMs:  $\gamma'=S\gamma S^{\rm T}$. The singular value decomposition
of
a real symplectic matrix gives \cite{dut}:
\begin{equation}\label{eu}
  S=K\Big[\bigoplus_{i=1}^{N} S(r_i) \Big] L,
\end{equation}
where $S(r_i)$ is a one-mode squeezing matrix (symplectic and nonorthogonal) with
$r_i$ the squeezing parameter:
\begin{equation}\label{sqr}
  S(r_i)=\left({\begin{array}{*{20}c} {\rm e}^{-r_{i}} & 0  \\
0 & {\rm e}^{r_{i}} \end{array}}\right),
\end{equation}
and $K$, $L$ are symplectic and orthogonal matrices. Denote by $K(2N)$ the group
of
orthogonal symplectic matrices isomorphic to the group of complex unitary
matrices
$U(N)$ \cite{dut}:
 \begin{equation}\label{ku}
    K(2N)=\{S(X,Y)|X-{\rm i}Y\in U(N)\},
  \end{equation}
  where the corresponding symplectic matrices are given by:
 \begin{equation}\label{sorth}
   S(X,Y)=\left({\begin{array}{*{20}c} X & Y  \\
-Y & X
\end{array}}\right).
\end{equation}

According to the Williamson theorem \cite{will} every real symmetric matrix
$M\geq0$
can be brought to a diagonal form through symplectic transformations as follows:
\begin{equation}
SMS^T={\rm diag}(s_1, s_1,\dotsc, s_N, s_N),
\end{equation}
where $s_1,\dotsc,s_N\geq 0$ are called symplectic eigenvalues of $M$.  By
\begin{equation}\label{defstr}
{\rm str}[M]:=\sum_{i=1}^N s_i
\end{equation}
we denote the symplectic trace of the matrix $M$.

\section{Gaussian quantum steering}
Consider the situation where Alice and Bob are two distant parties sharing a
common
state $\rho _{AB}$,  and Alice  performs local measurements $\hat A$  on her
state,
with eigenvalues $a$.  A formal definition of steering says that Alice is not
able
to steer Bob's state if there exists an ensemble $F=\{p_{\eta} \rho_{\eta}\}$ of
preexisting local hidden states $\rho_{\eta}$ with probabilities $p_\eta$, and a
stochastic map $P(a|\hat A,
\eta)$\footnote{$P(a|\hat A, \eta)$ denotes the probability distribution of Alice
to
obtain outcome $a$ when measuring $\hat A$ and given a local hidden variable
$\eta$.} from the hidden variable $\eta $ to $a$,  such that Bob's conditioned
state
after Alice's measurement is given by \cite{wis}:
\begin{equation}\label{rosteer}
  \tilde{\rho}_B^A=\sum_{\eta}P(a|\hat A, \eta) \rho_{\eta} p_{\eta}.
\end{equation}
This represents the local hidden state (LHS) model, where Bob checks if Alice can
simulate the state $\tilde{\rho}_B^A$ based on her knowledge of the parameter
$\eta$, by drawing the states $\rho_\eta$ according to the distribution $p_\eta$.
Conversely, if Bob cannot find any ensemble $F$ and map $P(a|\hat A, \eta)$
satisfying Eq. (\ref{rosteer}), then Bob must admit that Alice can steer his
system.
\\
\indent In the Gaussian realm where the joint state $\rho_{AB}$ is a Gaussian
state
with CM $\gamma_{AB}$, we consider that Alice's measurement is also Gaussian
(i.e.
mapping Gaussian states into Gaussian states). As discussed in the previous
section a
Gaussian measurement is represented by a positive operator $\hat A$ with a
Gaussian
characteristic function as in Eq. (\ref{gchf}) with CM $T^A$, satisfying
$T^A+{\rm
i} \Omega_{N_A}\geq 0$. When Alice performs a measurement $\hat A$ and gets an
outcome $a$, Bob's conditioned state $\tilde{\rho}_B^A$ is a Gaussian state with
CM
given by $\gamma_B^A=\gamma_B-\gamma_{12}^{\rm T}(\gamma_A+T^A)^{-1}\gamma_{12}$
\cite{ser}. In Ref. \cite{wis} a criterion for Gaussian steerability of CM
$\gamma_{AB}$ was derived:\\

\textit{\textbf{Theorem 1.} \cite{wis} A Gaussian state with covariance matrix
$\gamma_{AB}$ with the block form defined in Eq. (\ref{bl}) is Alice $\to$ Bob
non-steerable by Gaussian measurements if and only if
\begin{equation}\label{scm}
  \gamma_{AB}+0_{A}\oplus {\rm i} \Omega_{N_B}\geq 0
\end{equation}
or, equivalently\footnote{This follows from applying the positivity conditions
from
Lemma 1 to the total matrix in Eq. (\ref{scm}).}
\begin{equation}\label{schu}
  \gamma_A\geq 0 \quad \text{and}\quad\gamma_{AB}/\gamma_A+{\rm i}\Omega_{N_B}
  \geq 0
\end{equation}
holds,  where $\gamma_{AB}/\gamma_A:=\gamma_B-\gamma_{12}^{\rm T}\gamma_A^{-1}
\gamma_{12}$ is the Schur complement of $\gamma_{AB}$ with respect to submatrix
$\gamma_A$ and $\Omega_{N_B}$ is the symplectic matrix of $N_B$ modes as defined
in
Eq. (\ref{sym}).}\\

From the previous section we know that the Schur complement of a CM is a positive
matrix.
However, for a non-steerable Alice to Bob Gaussian state the Schur complement of
its CM
with respect to $\gamma_A$ has to satisfy a stronger condition than positivity,
namely the CMC from Eq. (\ref{unc}). An important result of this article is given
by
the following theorem on Gaussian non-steerability. \\

\textit{\textbf{Theorem 2.}  A bipartite quantum Gaussian state $\rho_{AB}$ with
covariance matrix $\gamma_{AB}$ with blocks defined in Eq. (\ref{bl}) is
Alice $\to$ Bob non-steerable by means of Gaussian measurements if and only if
there
exists a covariance matrix corresponding to Bob's system $\sigma_B$ satisfying
$\sigma_B+{\rm i}\Omega_{N_B}\geq 0$, such that:
\begin{equation}\label{cmc}
  \gamma_{AB}\geq 0_A \oplus \sigma_B.
\end{equation}}\vspace{0.03mm}
\begin{proof}
$\Rightarrow$ Theorem 1 states that the Schur complement $\gamma_{AB}/\gamma_A$
of a
non-steerable CM $\gamma_{AB}$ is also a CM, i.e.
$\gamma_{AB}/\gamma_A+{\rm i}\Omega_{N_B} \geq 0$. Based also on the definition
of the Schur
complement in Eq. (\ref{scb}) we state that if a CM $\gamma_{AB}$ is
non-steerable
then there exists a CM $\sigma_B$ fulfilling Eq. (\ref{cmc}), and it is the Schur
complement $\gamma_{AB}/\gamma_A$.\\
$\Leftarrow$  Conversely, if the relation in Eq. (\ref{cmc}) is fulfilled for
some CM $\sigma_B$, satisfying CMC $\sigma_B+{\rm i}\Omega_{N_B} \geq 0$, then by
the definition of the Schur complement in Eq. (\ref{scb}) it follows
that there exists a positive semi-definite matrix $P\geq 0$ such that:
$\sigma_B+P=\gamma_{AB}/\gamma_A$. Therefore, the Schur complement
$\gamma_{AB}/\gamma_A$ also has to fulfill the CMC, since $\sigma_B+P+{\rm
i}\Omega_{N_B} \geq P\geq 0$, which based on Theorem 1, means that
$\rho_{AB}$ is a non-steerable quantum state.
\end{proof}
 In Ref. \cite{ji} a one way proof of this theorem based on local
 uncertainty relations (LURs) is provided. The non-steerability criterion
 formulated in
 Theorem 2 shows a strong similarity to the separability of Gaussian states.  A
 continuous variable state with CM $\gamma$ is separable with respect to parties
 $A, B$ if and only if there exist local CMs $\gamma_A$ and $\gamma_B$ for each
 partition, such that $\gamma\geq \gamma_A \oplus \gamma_B$ \cite{wolf}. Hence,
 it
 is obvious that any separable Gaussian state is also non-steerable by Gaussian
 measurements since $\gamma_A\oplus\gamma_B\geq 0_A\oplus \gamma_B$ holds. Thus,
 Gaussian non-steerability represents a stronger condition than separability of
 Gaussian states.

\section{Steering witnesses based on second moments}
Based on Theorem 2 we can define the set of non-steerable CMs as follows:
\begin{equation}\label{nons}
\Gamma_{A\not\to B}(\mathbb{R}^{2N}):=\{\gamma_{AB}|~ \gamma_{AB}=0_A\oplus
\sigma_B+P, ~\text{with} ~\sigma_B+{\rm i}\Omega_{N_B}\geq 0~ \text{and}~ P\geq 0
\},
\end{equation}
where $A\not \to B$ denotes Alice to Bob non-steerability, and $P$ is a
positive-semidefinite matrix. The set of non-steerable CMs by Gaussian
measurements forms a closed convex
subset of the space of all covariance matrices, similarly to the set of all CMs
and the set of separable CMs. This allows  to completely describe
the set of non-steerable CMs by a family of linear inequalities representing the
steering witnesses (SWs).

\textit{\textbf{Definition 1.} We define the set of real symmetric $2N\times
2N$-matrices
\begin{equation}\label{witdef}
  \mathcal{Z}_{A\not \to B}(\mathbb{R}^{2N})=\{Z|Z\geq 0, \forall \gamma_{AB}\in
  \Gamma_{A\not \to B}(\mathbb{R}^{2N}):{\rm Tr}[Z\gamma_{AB}]\geq 1 \},
\end{equation}
where $\gamma_{AB}$ is the CM of a bipartite system with $N=N_A+N_B$ modes,
$A\not
\to B$ denotes Alice to Bob non-steerability, where $\Gamma_{A\not\to
B}(\mathbb{R}^{2N})$ is defined in Eq.(\ref{nons}). All matrices
$Z\in\mathcal{Z}_{A\not \to B}(\mathbb{R}^{2N})$ for which there exists $\gamma$
with ${\rm Tr}[Z\gamma]<1$ will
be called steering witnesses (SWs).}\\

This Definition 1 is very similar to the Definition 6.4 in Ref. \cite{janet} for
entanglement witnesses (EWs) since it relies solely on the fact that the set of
non-steerable CMs and the set of separable CMs are both convex and closed sets.
There it is also shown that for any such hyperplane $Z\geq 0$ holds, which
follows from the positivity of CMs. We proceed, in analogy with the methods used
in Ref. \cite{janet}, to characterise the set of SWs and prove some of their
properties.\\

\textit{\textbf{Theorem 3.} Let $Z$ be a real symmetric $2N\times 2N$ matrix on a
phase space of $N=N_A+N_B$ modes,  with $Z_A$ and $Z_B$ denoting the block
diagonal
submatrices of $Z$ corresponding to subsystems of Alice and Bob, respectively.
Then
$Z$ is a steering witness, namely $Z\in \mathcal{Z}_{A\not \to
B}(\mathbb{R}^{2N})$,
if and only if
\begin{align}
Z&\geq 0,\label{cond1} \\
{\rm str}[Z_B]&\geq \frac{1}{2},\label{cond2}\\
{\rm str}[Z]&<\frac{1}{2}.\label{cond3}
\end{align}
}\color{black}
\begin{proof}  Below Theorem 6.2 in Ref. \cite{janet} it is proven that any
hyperplane cutting the set of CMs is represented by a positive semi-definite
matrix,
and therefore, condition $Z\geq 0$ holds for any EW and SW as well. In addition,
in
Theorem 6.3 in the same reference it is proven that such a test represented by
$Z\geq
0$ and satisfying also ${\rm str}[Z]\geq \frac{1}{2}$ is equivalent to obtaining
${\rm Tr}[Z \gamma]\geq 1$ for all CMs $\gamma$. In the following we will prove
that
condition ${\rm str}[Z_B]\geq \frac{1}{2}$ guarantees that  ${\rm
Tr}[Z\gamma]\geq1$ for all $A \rightarrow B$ non-steerable CMs. Therefore,
condition
${\rm str}[Z]<\frac{1}{2}$ together with conditions (\ref{cond1}, \ref{cond2})
assure that there exists a CM $\gamma$ such that ${\rm Tr}[Z\gamma]<1$, i.e. it
is $A \rightarrow B$ steerable. \\
Consider the SWs based on second moments of the following block form:
\begin{equation}\label{wit}
  Z=\begin{pmatrix}
      Z_A & Z_{12} \\
      Z_{12}^{{\rm T}} & Z_B
    \end{pmatrix}.
\end{equation}
In the following we will use an important result discussed in Ref. \cite{janet}
\footnote{The expression $2\,{\rm str}[Z]$ represents the value of the ground
state
energy of the Hamiltonian defined as $\hat H=Z\sum_{kl} \hat R_k \hat R_l$, where
$\mathbf{\hat R}=(\hat x_1,\hat p_1,...,\hat x_N,\hat p_N)^{{\rm T}}$ is the
vector of
canonical operators.}:
\begin{equation}\label{min}
  \min_{\sigma_B} {\rm Tr}[Z_B \sigma_B]=2\,{\rm str}[Z_B],
\end{equation}
where the minimization is performed over the set of CMs $\sigma_B$.\\
$\Rightarrow$ We have to prove that condition in Eq.(\ref{cond2}) holds for any
SW
described in Definition 1 (\ref{witdef}). Based on Theorem 2 for any
$\gamma_{AB}\in \Gamma_{A\not \to B}(\mathbb{R}^{2N})$ there exists a CM
corresponding to the quantum state of Bob $\sigma_B$ such that:
\begin{equation}\label{pp}
  {\rm Tr}[Z \gamma_{AB}]\geq  {\rm Tr}[Z (0_A\oplus \sigma_B)]
\end{equation}
holds. Moreover, the matrix $0_A\oplus \sigma_B$ is a non-steerable CM itself,
and
from the Definition 1 of SWs in Eq. (\ref{witdef}) we have the following
condition:
\begin{equation}\label{ppm}
    {\rm Tr}[Z (0_A\oplus \sigma_B)]={\rm Tr}[Z_B \sigma_B]\geq 1.
\end{equation}
Using Eq. (\ref{min}) we conclude that SWs have to fulfill the condition
\begin{equation}\label{pl}
  {\rm str}[Z_B]\geq \frac{1}{2}.
\end{equation}

$\Leftarrow$ Conversely, we prove that a matrix $Z$ satisfying the conditions in
Eq.
(\ref{cond2}) represents a SW with ${\rm Tr}[Z\gamma_{AB}]\geq1$ for any
non-steerable CM $\gamma_{AB}\in\Gamma_{A\not\to B}(\mathbb{R}^{2N})$. Starting
from
Eq. (\ref{cond2}) and using Eq. (\ref{min}) we have that
\begin{equation}\label{pml}
  {\rm Tr}[Z_B \sigma_B]\geq 1
\end{equation}
holds for any valid CM $\sigma_B$. Now, any non-steerable CM $\gamma_{AB}$
fulfills
$\gamma_{AB}\geq 0_A\oplus\sigma_B$ (Theorem 2)
such that the condition ${\rm Tr}[Z\gamma_{AB}]\geq {\rm Tr}[Z_B \sigma_B]$
holds.
Therefore we get for any non-steerable state ${\rm Tr}[Z \gamma_{AB}]\geq 1$.
\end{proof}
\vspace*{0.3cm}
We are now in the position to summarize the results for the detection of Gaussian
steering with witnesses based on second moments. \\

\textit{\textbf{Theorem 4. [Steerability]} A CM $\gamma_{AB}$ of two parties
consisting of
$N=N_A+N_B$ modes is Alice to Bob steerable by means of Gaussian measurements if
and
only if there exists a $Z$  such that:
\begin{equation}
  {\rm Tr}[Z\gamma_{AB}] < 1,
  \end{equation}
  where $Z$ is a real symmetric $2N\times 2N$ matrix satisfying
  \begin{equation}\label{witstr}
 Z \geq 0 \quad  and \quad {\rm str}[Z_B]  \geq\frac 1 2,
\end{equation}
where $Z_B$ denotes the principal submatrix of $Z$ belonging to the subsystem of
Bob. Matrices $Z$ are called steering witnesses based on second moments.}\\

The EWs defined in Refs. \cite{hyll,janet} that detect bipartite entanglement
fulfill the condition ${\rm str}[Z_A]+{\rm str}[Z_B]\geq \frac{1}{2}$, where
$A, B$ denote the two parties, and $Z_A, Z_B$ are the block diagonal matrices of
the
EW associated with each party. SWs are characterised by stronger constraints than
these and
therefore, any SW based on second moments is also an EW. This is consistent with
the
 fact that any steerable
state necessarily contains also entanglement.

\section{Linear constraints for steering witnesses}

\indent In the following we present linear conditions for the SWs that are
stronger than Eq. (\ref{witstr}) in Theorem 4, such that the SWs can be
calculated by a semidefinite program (SDP) to detect
steering in a given CM. An analogous idea for EWs was developed in Ref.
\cite{tamih}.\\

\textbf{Proposition 1.} \textit{For the steering witness $Z$ of an $N-$mode
covariance
matrix steerable from Alice to Bob, with $N=N_A+N_B$, the inequalities
(\ref{witstr}) are satisfied if (if and only if for $N_B=1$) the following
conditions are fulfilled:
\begin{eqnarray}\label{prop}
& Z\geq 0, \nonumber\\
& Z_B+{\rm i} \frac{1}{2N_B}\Omega_{N_B} \geq 0.
\end{eqnarray}}
\begin{proof}
Since $Z$ is a positive semidefinite matrix, also the principal submatrix $Z_B$
is
positive semidefinite. For every such matrix there exists a symplectic
transformation $S$ that brings it to its Williamson normal form as follows
\footnote{Given that ${\rm Tr}[M]\geq 2  \ {\rm str}[M]$ holds for any positive
matrix $M$ \cite{bathia}, the symplectic transformations do not only preserve the
symplectic eigenvalues, but also positivity.}:
\begin{equation}\label{poz}
S(Z_B+{\rm i} \frac{1}{2N_B} \Omega_{N_B} )S^{T}=Z_B^w+{\rm i}
\frac{1}{2N_B}\Omega_{N_B},
\end{equation}
with $Z_B^w={\rm diag}(z_1,z_1,\cdots, z_{N_B},z_{N_B})$, where $z_j$,
$j=1,\cdots
N_B$ are positive symplectic eigenvalues of $Z_B$. By imposing the positivity
condition on the eigenvalues of the total matrix in Eq. (\ref{poz}) we arrive at
the
following inequality for the symplectic eigenvalues:
\begin{equation}\label{zip}
  z_j\geq \frac{1}{2 N_B}, \quad j=1,\cdots,N_B.
\end{equation}
Now, the sum of symplectic eigenvalues gives:
\begin{equation}
\sum_{j=1}^{N_B}z_j\geq \sum_{j=1}^{N_B}\frac{1}{2 N_B}=\frac 1 2.
\end{equation}
\end{proof}

\indent Relation (\ref{zip}) explains why the conditions on SWs given in the
Proposition 1
are stronger than Eq. (\ref{witstr}) from Theorem 4: it imposes a lower bound on
the
symplectic eigenvalues that is strictly greater than zero, as $\frac{1}{2
N_B}>0$. This
assures that the sum of symplectic eigenvalues exceeds $\frac{1}{2}$, however, it
rules out the possibility that some symplectic eigenvalues may satisfy
$z_j\in(0,\frac{1}{2 N_B})$. An exception represents the case when the
steered party has one mode ($N_B$=1) and $Z_B$  in Eq. (\ref{prop}) is a $2\times
2$ matrix with only one symplectic eigenvalue satisfying $z\geq \frac{1}{2}$,
which is equivalent to relation (\ref{witstr}). However, the interval of
forbidden symplectic eigenvalues $z_j\notin(0,\frac{1}{2 N_B})$ is the largest
for $N_B=2$, while in the limit of large  $N_B\rightarrow\infty$ we have
$\frac{1}{2 N_B}\rightarrow 0$, and the constraints in Proposition 1 become
equivalent to the constraints fully characterizing SWs in Theorem 4.

\subsection{Constructing witnesses}
We are now able to define an SDP by constructing the witnesses from given
(random) measurements using the linear constraints in the Proposition 1. Given
the
repeated independent measurements $P_j$ on the CM, the witness operator is
represented by $Z=\sum_{j} c_j P_j$, where the coefficients $c_j$ are the output
of
the optimization algorithm. According to the Proposition we can detect the
Gaussian
steering from Alice to Bob in two-mode CMs with the following SDP:
\begin{mini}
{}{\bf{c\cdot m}}{}{}
\addConstraint{Z=\sum_j c_j P_j}
\addConstraint{ Z=\left({\begin{array}{*{20}c}
      Z_A & Z_C \\
      Z_C^{\rm T} & Z_B \end{array}}\right)\geq 0 }
\addConstraint{Z_B+{\rm i} \frac{1}{2} \Omega_{N_B} \geq 0 },\label{min2}
\end{mini}
where $\bf{c}$ is the vector of coefficients $c_j$ and $\bf{m}={\rm
Tr}[P\gamma]$,
with $\bf{P}$ being the vector of measurement matrices $P_j$, which represent the
homodyne detection measurements as constructed in Ref. \cite{tamih}.

Consider an experimental scheme detecting two-mode CMs using a single homodyne
detector, a phase shifter of angle $\varphi$ between the vertical and horizontal
polarization components, a polarization rotator on angle $\phi$ and a polarizing
beam splitter, for mixing the initial modes denoted by $\hat a$ and $\hat b$
\cite{dauria}. The mode $\hat k$ arriving at the detector has the following
expression:
\begin{equation}\label{kk}
  \hat k=\exp({\rm i} \varphi) \cos \phi \ \hat a +\sin \phi \ \hat b.
\end{equation}
The generalized quadrature operator is given by:
\begin{equation}
  \hat x_{\theta}=\frac{\exp{(-{\rm i}\theta)}\hat k+\exp{({\rm i}\theta)}\hat
  k^{\dag}}{\sqrt{2}},
\end{equation}
which covers the whole continuum of quadratures in terms of initial modes $\hat
a$
and $\hat b$ for $\theta \in [0, \pi]$,  $\phi\in [0,\pi]$ and $\varphi\in[0,2
\pi)$.
With homodyne detection one can measure the statistical moments of the quadrature
up
to the second order \cite{dauria}, and therefore, one can readily calculate the
variance of the quadrature as follows:
\begin{equation}\label{p}
\langle \hat x_{\theta}^2 \rangle-\langle \hat x_{\theta}\rangle^2=\Tr[P \gamma],
\end{equation}
where $P$ is the measurement matrix of the variances we use in Eq. (\ref{min2}):
\begin{equation}
 P=u u^{\rm T}, \quad u^{\rm T}=\begin{pmatrix}\cos \phi \cos(\theta-\varphi)&
 \cos
 \phi \sin(\theta-\varphi)&\sin \phi \cos \theta & \sin \phi \sin \theta
 \end{pmatrix}.
\end{equation}
The $N-$mode CM is a symmetric, real $2N\times2N$ matrix with $N(2N+1)$
independent
parameters. Therefore, for the CM reconstruction of a two-mode Gaussian state one
needs $10$ distinct measurement directions given by the angles $\theta$, $\phi$
and
$\varphi$. This scheme can be extended to $N$ modes with a single
homodyne detector by using the same two-mode combination scheme $N-1$ times.
Starting with three initial modes $\hat a$, $\hat b$ and $\hat c$, the
generalized mode arriving at the detector would be:
\begin{equation}
\hat k= \exp({\rm i} \varphi_1) \cos \phi \  \hat a + \exp({\rm i} \varphi_2)\sin
\phi \cos \psi \ \hat b+ \sin \phi \sin \psi \  \hat c.
\end{equation}

\section{Detection of Gaussian steering}
Our method of detection relies on the SDP described in Eq. (\ref{min2}), where we
run the optimisation starting with two measurement settings, and then adding
other
measurements, one by one, until steering is detected. In this way we are able
to evaluate the efficiency of steering detection in terms of the number of
required
measurement settings. In the case of general quantum states of
unknown origin the best strategy could be to perform random measurements, which
in
our case reduces to choosing random values for angles $\theta$, $\phi$ and
$\varphi$.\\
\indent A measure for quantifying steering of a bipartite Gaussian
state with CM defined in Eq. (\ref{bl}) was proposed in Ref.
\cite{kog}.  In particular, the
nonsteerability condition in Eq. (\ref{schu}) is equivalent to $\tilde \mu_j\geq
1$
for all $j=1, \dots, N_B$, where $\tilde \mu_j$ represent the symplectic
eigenvalues
of the Schur complement $\gamma_{AB}/\gamma_A$.  The Gaussian $A\rightarrow B$
steerability can then be quantified via \cite{kog}:
\begin{equation}\label{kog1}
\mathcal{G}^{A\to B}(\gamma_{AB})=\max\bigg\{0, -\sum_{j:\tilde \mu_j<1} \ln
(\tilde
\mu_j) \bigg\}.
\end{equation}
This expression is invariant under local symplectic transformations at the CM
level, and
has a formal similarity to the logarithmic negativity \cite{plen}, which for
Gaussian states
quantifies how much the partially transposed CM fails to fulfill CMC in Eq.
(\ref{cmc}). The quantification of Gaussian steerability takes a simpler form
when the
steered party, e.g. Bob, has one mode only $(N_B=1)$ \cite{kog}:
\begin{equation}\label{kogeq}
\mathcal{G}^{A\to B}(\gamma_{AB})=\max\{0,\frac{1}{2}\ln \frac{\det
\gamma_A}{\det\gamma_{AB}}\}.
\end{equation}
Steering can also be quantified by a minimal SW, where $Z_{min}$ corresponds to
the smallest possible
value ${\rm Tr}[Z_{min} \gamma_{AB}]= w_{min}$ for a CM $\gamma_{AB}$. The
program
proposed in Eq. (\ref{min2}) may reach the minimal value $w_{min}$ with more
measurements than in full tomography, since the linear constraints used in the
optimization for finding the appropriate SWs are stronger than required in
Theorem 4. For two-mode CMs $\gamma_{AB}$ our method shows
that the minimal SW $Z_{min}$  is related to the steerability measure from Ref.
\cite{kog} as follows:
\begin{equation}\label{rel}
  \mathcal{G}^{A\to B}(\gamma_{AB})=\ln{\frac{1}{{\rm Tr}[Z_{min} \gamma_{AB}]}}.
\end{equation}

\subsection{Two-mode squeezed vacuum states}
First we test our method on the class of two-mode squeezed vacuum states (SVS),
which are easily accessible in experiments \cite{weed}. The CM is of the
following
form:
\begin{equation}\label{sts}
\gamma=\left({\begin{array}{*{20}c}
\cosh 2r & 0 &  \sinh 2r & 0 \cr
0 & \cosh 2r & 0 & - \sinh 2r \cr
 \sinh 2r & 0 & \cosh 2r & 0 \cr
0 & - \sinh 2r & 0 & \cosh 2r
\end{array}}\right),
\end{equation}
where $r$ is the squeezing parameter. For this particular case, using formula
(\ref{kogeq}), the expression quantifying steering is a monotonic function in
$r$:
\begin{equation}\label{stsvs}
\mathcal{G}^{A\to B}(\gamma_{sv})= \ln{\cosh^2{(2 r)}}.
\end{equation}
\indent In Fig. 1 we show the fraction of SVS certified with steering with
respect
to the number of measurements performed randomly. First, we notice the similarity
between the detection of SVS with higher steering and the detection of high
entanglement \cite{tamih}, since both require $8-9$ measurements on average.
However, SVS with less entanglement can be certified in many cases
with $6-7$ measurements, on the other side the detection of low steering mostly
requires $10$
measurements (full tomography). The demand on more measurements for steering
detection in comparison with entanglement is consistent with the fact that SWs
based
on second moments satisfy stronger constraints than EWs (see discussion below
Theorem 4).

\begin{figure}[h]
\centering
\includegraphics[width=0.95\columnwidth]{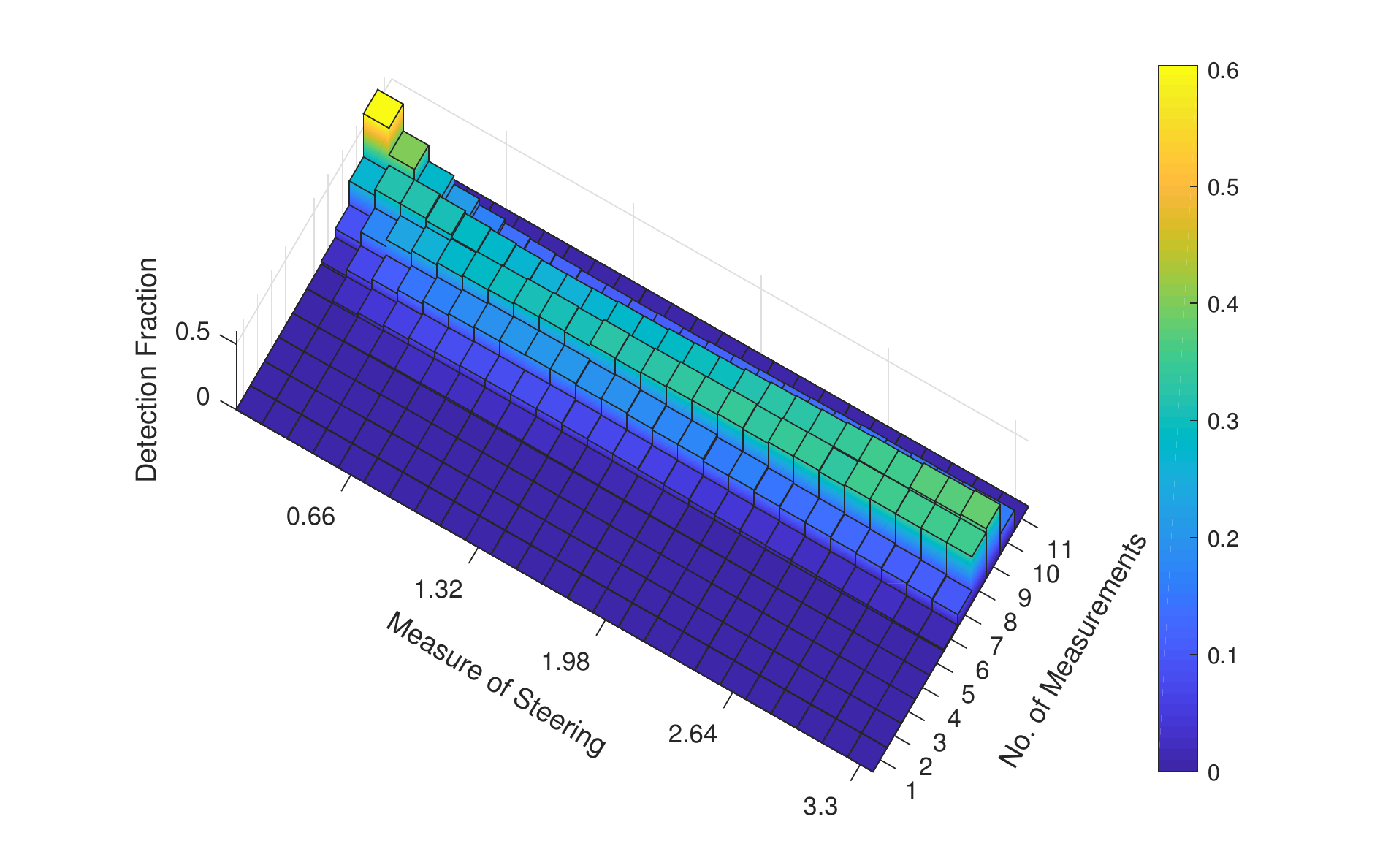}
\caption{Fraction of steering detection for two-mode squeezed vacuum states with
$r\in[0,2]$, see  Eq. (\ref{sts}): $5\times10^5$ samples. The data  are
normalized such that they sum up to $1$ for each column. The measure of Gaussian
Steering $ ( \mathcal{G}^{A\to B}(\gamma_{AB}))$ is defined in Eqs. (\ref{kogeq})
and (\ref{stsvs}).}
\end{figure}

\subsection{Random two-mode covariance matrices}
For the case of general unknown CMs we test our method by generating random
two-mode
CMs and using random measurements. We start with a CM of a thermal state, which
is a
diagonal matrix with the symplectic eigenvalues $\nu_i\geq 1$  for every mode
$i=1,...,N$, related to the thermal photon number $n_i$ as $\nu_i=2 n_i+1$
\cite{weed}:
\begin{equation}\label{th}
\gamma_{th}= \bigoplus_{i=1}^N \left({\begin{array}{*{20}c} \nu_i & 0 \\ 0 &
\nu_i
\\ \end{array}}\right),
\end{equation}
where $\nu_i$ are randomly generated from a uniform distribution in a finite
interval $[1,t]$, $t>1$. Any general CM can be obtained from a thermal state CM
that
undergoes a symplectic transformation:
\begin{equation}\label{th1}
\gamma=S  \gamma_{th} S^{\rm T}.
\end{equation}
For random symplectic matrices we use the singular value decomposition (see Eq
(\ref{eu})). We
randomly generate unitary matrices $X$ and $Y$ from the Haar distribution, while
for
one-mode squeezers $S(r_i)$ we create random parameters $r_i$ by a uniform
distribution in a finite interval. The MATLAB code generating symplectic matrices
as
described above, was developed in Ref. \cite{jagger}.
\begin{figure}[h]
\centering
\includegraphics[width=0.95\columnwidth]{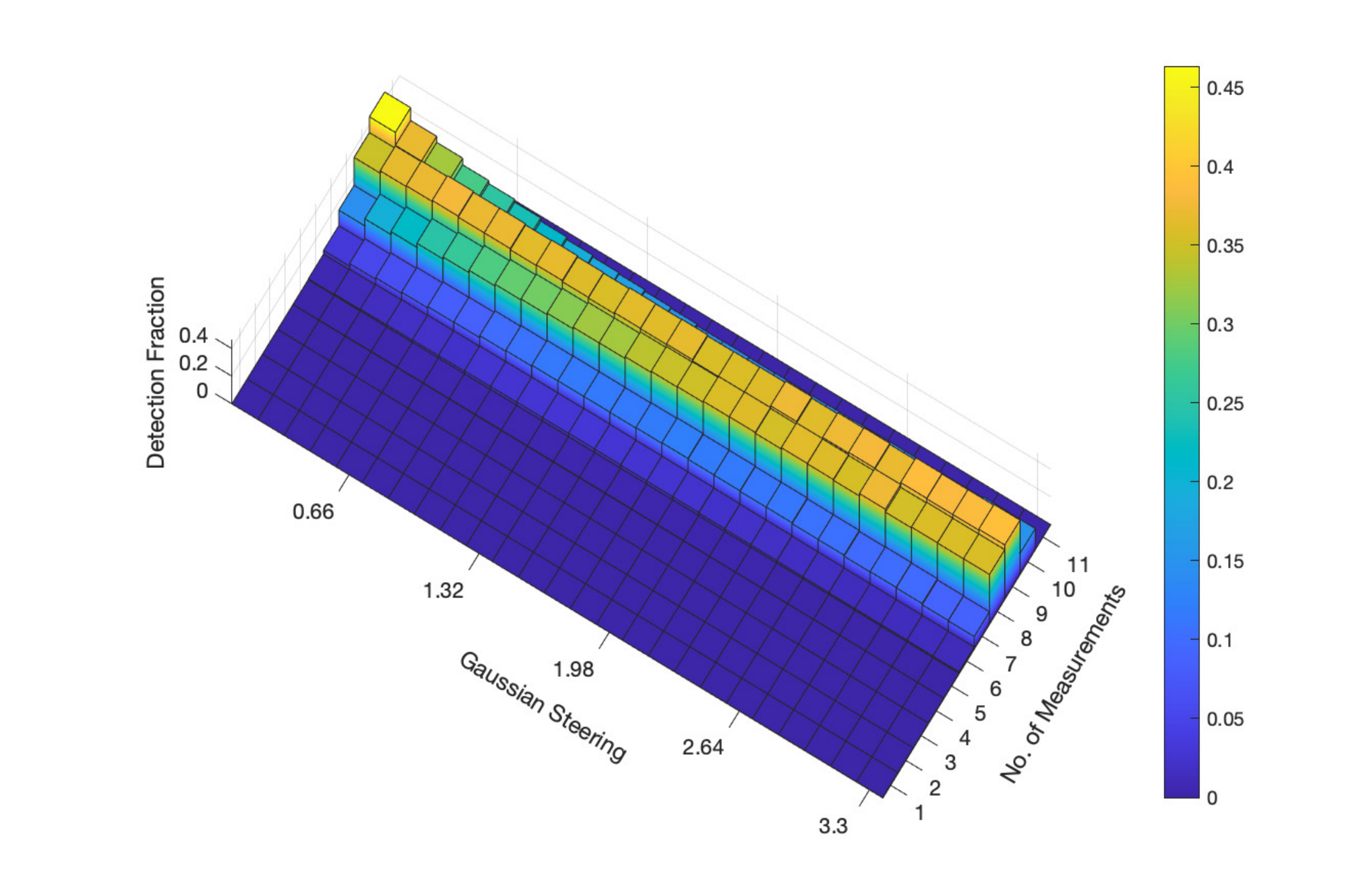}\\
\caption{Fraction of steering detection for two-mode randomly generated CMs of
thermal states with symplectic eigenvalues $\nu_i\in [0,5]$ and symplectic
transformations with squeezing parameter $r_i\in [0,2]$.  It represents $5\times
10^5$ runs of the algorithm where measurement directions are added successively
and the
SW is evaluated at every round until steering is certified.  The data  are
normalised
for every value of steering $(  \mathcal{G}^{A\to B}(\gamma_{AB}))$ such that
they sum up to 1.       }
\end{figure}

In Figure 2  we  illustrate the data from running the algorithm for $5\times
10^5$
different randomly generated CMs of  thermal states with symplectic eigenvalues
randomly generated from a finite interval $\nu_i\in[0,5]$ and symplectic
eigenvalues
with squeezing parameters $r_i\in [0,2]$.  It shows the efficiency of steering
detection by calculating the SW (see Eq. (\ref{min2})) for every newly added
measurement direction
until it finds steering.  Comparing to the particular case of SVS discussed in
Figure 1 the detection of steering in general CMs shows slight improvement for
low
steering where only  for  a fraction of $0.45$ of the CMs steering was certified
by
the 10th measurement.  For high steering one may need most of the time $8 - 9$
measurements for the detection of steering.  The detection of
two-mode Gaussian steering shows a similar behaviour as the detection of
entanglement in randomly generated CMs \cite{tamih}. The stronger the
correlation the easier steering or entanglement can be detected ( i.e. it
requires fewer measurements).

\subsection{Three-mode continuous variable GHZ states}
The CV counterpart for the GHZ three qubit states is experimentally created from
three squeezed beams (two position-squeezed beams with squeezing $r_p$ and one
momentum-squeezed beam with squeezing $r_m$) which are mixed in a double beam
splitter \cite{lock}. The CM obtained in this manner corresponds to a pure,
symmetric, genuine  multipartite  entangled three-mode Gaussian state, also known
as CV GHZ state \cite{addss}:
\begin{equation}\label{ghz}
  \gamma_{GHZ}=\begin{pmatrix}
           a & 0 & c & 0 & c & 0 \\
           0 & b & 0 & -c & 0 & -c \\
           c & 0 & a & 0 & c & 0 \\
           0 & -c & 0 & b & 0 & -c \\
           c & 0 & c & 0 & a & 0 \\
           0 & -c & 0 & -c & 0 & b
         \end{pmatrix},
\end{equation}
where $a$ is related to the squeezing parameters in momentum $r_m$ and position
$r_p$ as \cite{addss}:
\begin{equation}
a=\frac{1}{3}\sqrt{4\cosh{2(r_m+r_p)}+5},
\end{equation}
 and
\begin{align}\label{cb}
  b & = \frac{1}{4}(5 a-\sqrt{9 a^2-8}),\\
  c & =\frac{1}{4}(a-\sqrt{9 a^2-8}).
\end{align}
In fact, a proper (unnormalized) CV GHZ state is a simultaneous eigenstate with
zero eigenvalues of total momentum $\hat p_1+\hat p_2+\hat p_3$ and relative
positions $\hat x_i-\hat x_j$ $(i,j=1,2,3)$, whereas the CV state described in
Eq. (\ref{ghz}) approaches the CV GHZ state in the limit of infinite squeezing
$(a\rightarrow\infty)$ \cite{lock2}.
\begin{figure}[h]
\centering
\includegraphics[width=1\columnwidth]{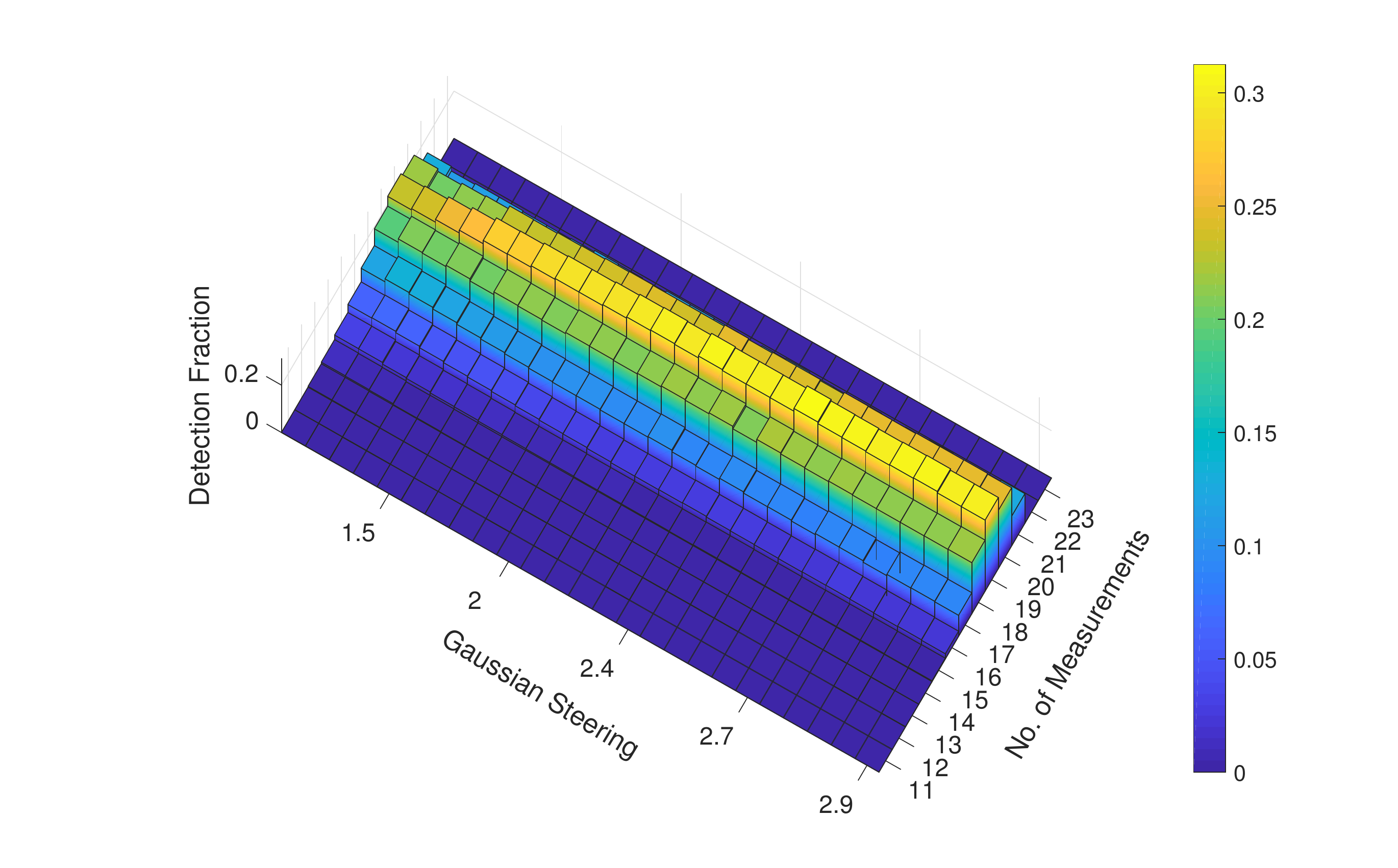}
\caption{ Fraction of steering detection for three-mode CV GHZ states with
$a=2,3,...,26$, see Eq. (\ref{ghz}). The data are obtained from $4.5\times10^5$
runs of the algorithm, and is normalized such that they sum up to 1. Steering is
quantified by $(  \mathcal{G}^{A\to B}(\gamma_{GHZ}))$ defined in Eq.
(\ref{kog1}).}
\end{figure}

\indent Let us consider the partition where Alice has one mode (it is not
important which mode, due to the special form of the CM in Eq. (\ref{ghz})) and
Bob has the other two modes. By calculating the symplectic eigenvalues of the
Schur complement corresponding to the situation where Alice is trying to steer
Bob's state by her choice of measurements (see Theorem 1), we obtain that the CV
GHZ state is steerable for $a>1$, and the amount of steering is given by the
measure in Eq. (\ref{kog1}).\\
\indent In Figure 3 we present  the efficiency of steering detection in GHZ
states described by the CM given in Eq. (\ref{ghz}) using the method of SWs as a
function of the number of random measurements. The algorithm was applied to
$4.5\times 10^5$ samples, where for each sample the number of measurement
settings to detect steering was recorded. For three-mode CMs there are $21$
independent measurement settings required for full tomography, whereas with our
method we detect steering mostly with $19$ measurements. This is despite the fact
that in our case $N_B=2$, and the linear constraints for the SWs defined in
Proposition 1 are stronger than required in Theorem 3. However, a small fraction
of $3\%$ of steerable GHZ states are detected by more measurements than needed in
full tomography.

\section{Statistical analysis}
In real experiments the homodyne data is obtained by $n$ repetitions of a
measurement with direction $\theta_i$, giving rise to a collection of outcomes
$X_{ij}=\langle \hat{x}_{\theta_i}\rangle_j$, ($j=1,\ldots,n$). In the case of
Gaussian states these outcomes are governed by the normal probability
distribution
$\mathcal{N}_i(\mu_i,m_i)$ with the mean $\mu_i=\langle
\hat{x}_{\theta_i}\rangle$,
and variance $m_i={\rm Tr}[P_i \gamma]=(\Delta \hat x_{\theta_i})^2$ (see Sec.
5).
The variances $m_i$ are estimated by the sample variances, which follow the
$\chi_{n_i-1}^2$ distribution \cite{chi}. From the statistical error carried out
by a $\chi_{n_i-1}^2$ distribution and using standard error propagation the
resulting
error for $\bar Z={\rm Tr[Z \gamma]}$ becomes \cite{tamih}:
\begin{equation}\label{indep}
  \Delta \bar Z=\sqrt{\sum_i \Big(\frac{d \bar Z}{d \bar P_i}\Big)^2 (\Delta \bar
  P_i)^2}=\sqrt{ \frac{2}{n-1} }\sqrt{\sum_i c_i^2 m_i^2}.
\end{equation}
\begin{figure}[h]
\centering
\includegraphics[width=0.8\columnwidth]{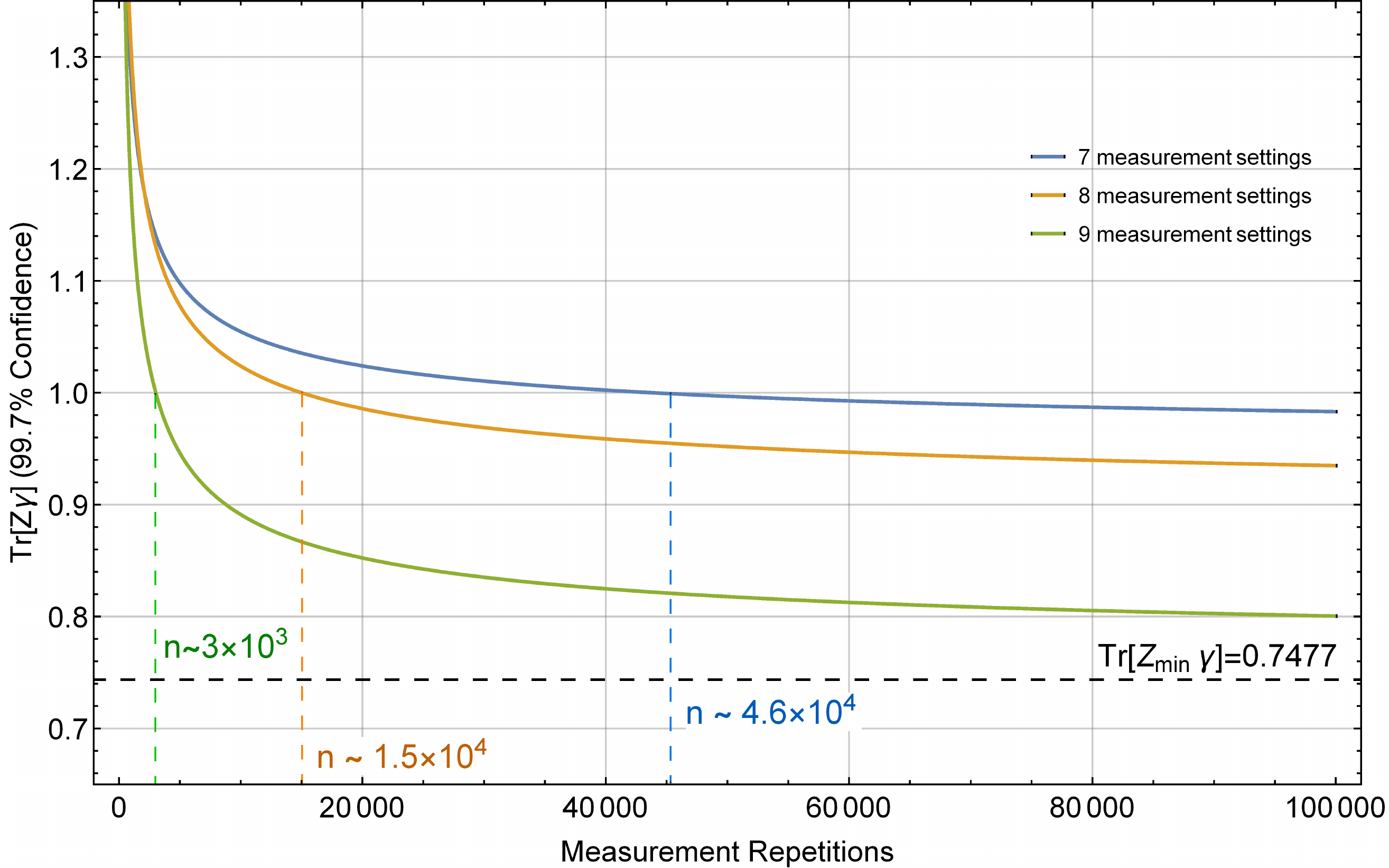}\\
\caption{The statistical estimate of $\bar Z$ for a squeezed vacuum state
covariance matrix $\gamma$
with the maximum of $3\sigma$ confidence interval according to Eq. (\ref{indep}).
The horizontal black dashed line indicates the minimal value of the witness for
the
considered CM $\Tr[Z_{min}\gamma]=0.7477$. The vertical dashed lines indicate the
number of measurement repetitions required to detect Gaussian steering with $7$
(blue), $8$ (orange) and $9$ (green) measurement settings.}
\end{figure}\\
\indent Our method of steering detection is based on the SDP presented in Section
5
where the coefficients $c_i$ are calculated using $m_i$ as input data. However,
the
formula in Eq. (\ref{indep}) does not take into account that these two variables
are
not independent. We overcome this difficulty by using two sets of homodyne data,
from the first one we derive the coefficients $c_i$ and the second one is used to
evaluate $\bar Z$ and $\Delta \bar Z$ \cite{mor}. \\
\indent In Figure 4 we show the estimated value of the SW $\bar Z$ with the $3
\sigma-$confidence for a steerable squeezed vacuum state CM $\gamma$  with ${\rm
Tr}[Z_{min} \gamma]=0.7477$, as a
function of measurement repetitions $n$. We can see that detecting steering with
$7-8$ measurement settings would require much more repetitions of the
measurements compared
to the case of $9$ measurement settings. As a result, the total number of
measurements is
one order of magnitude less for $9$ measurement settings ($\approx 3\times 10^3$)
than in  the case of $7$ different measurement settings ($\approx 4.6\times
10^4$).  Therefore, statistical analysis may help the experimentalist to decide
whether it is more advantageous to add new measurement directions or to increase
the number of repetitions of the measurements in order to detect
steering.

\section{Summary and conclusions}

\indent Non-steerable Gaussian states have  covariance matrices (CMs)
forming a closed and convex set, which gives rise to hyperplanes or steering
witnesses (SWs) , separating
any other point from this set.  This is easily seen from Theorem 2 which
provides a criterion for  non-steerable CMs.  Based on this result we
characterized the SWs for
Gaussian states by providing a set of constraints to be satisfied by any such
witness.

The method of detecting quantum correlation by witnesses has proven to be very
useful and accessible in experiments with entangled states \cite{toth}, while an
extensive study about the efficiency of entanglement detection with random
measurements in continuous variable states was presented in Ref. \cite{tamih}.
In
this article we have developed similar optimization tools for detecting steering
in Gaussian states with similar complexity and scaling behaviour.

We proposed a set of stronger linear constraints on  steering  witness operators
in comparison to  entanglement witnesses (EWs)  and studied
the efficiency of steering detection with respect to the number of measurement
settings. These linear constraints are proven to fully characterize the
set of steering witnesses when the steered party has only one bosonic mode. The
SDP we developed in Section 5 uses random measurements of
variances from the homodyne detection, as the building blocks for constructing
the
optimal steering test for a given unknown state.

In the case of two-mode squeezed vacuum states we noticed that detection of
steering
in Gaussian states requires more measurements on average than for entanglement
detection \cite{tamih}.   This lies in agreement with the fact that SWs satisfy
stronger conditions than EWs (see the discussion above Theorem 4).

We applied our method also for general random two-mode Gaussian states.  In this
case,  the
detection of both types of quantum correlations,  i. e.  entanglement and
steering,
has a behaviour confirming the general idea that higher quantum correlations are
easier to detect.  In particular,  high steering and entanglement in two-mode
states
are typically detected by $8-9$ measurement settings using our method,  while
full
tomography requires $10$ measurement settings. \\
\indent In addition, we provided an example of steering detection when the
steered party consists of more than one mode, namely the three-mode continuous
variable GHZ states. In this case the linear constraints used in the optimization
are stronger compared to the SW constraints in Theorem 4, reducing the set of
possible SWs. Nevertheless, the result shows that in most of the cases the GHZ
states are detected with steering by $19$ measurements or less, which is two
measurements fewer than in full tomography.  For $N$-mode CMs when the steered
party (Bob) has $N_B\rightarrow\infty$ number of modes we have shown that the
linear constraints tend to be equivalent to the exact constraints fully
characterizing the set of SWs. Therefore, our method of steering detection may
become better for larger number of modes. We provided also a statistical analysis
of our
method showing a good robustness to statistical errors.

\paragraph{Acknowledgements.}
 D.B. and H.K. acknowledge financial support by the QuantERA project QuICHE
via the German Ministry of
Education and Research (BMBF Grant No. 16KIS1119K) and by the Deutsche
Forschungsgemeinschaft (DFG, German Research Foundation) under Germany’s
Excellence Strategy - Cluster of Excellence Matter and Light for Quantum
Computing
(ML4Q) EXC 2004/1 - 390534769.  T.M. and A.I. acknowledge financial support
received from the Romanian Ministry of Research, Innovation and Digitization,
through the Project PN 23 21 01 01/2023.


\section*{References}
\begin{thebibliography}{110}
\bibitem{brus} D. Bru{\ss}, G. Leuchs, Eds., \emph{Quantum Information: From
    Foundations to Quantum Technology Applications, 2nd Edition}, Wiley-VCH
    (2019)
\bibitem{gras} F. Grasselli, G. Murta, H. Kampermann, D. Bru{\ss}, \emph{Entropy
    Bounds for Multiparty Device-Independent Cryptography}, PRX Quantum
    \textbf{2},
    010308 (2021)
\bibitem{metr} E. Polino, M. Valeri, N. Spagnolo,  and F. Sciarrino,
    \emph{Photonic quantum metrology}, AVS Quantum Sci. \textbf{2}, 024703 (2020)
\bibitem{schr1} E. Schr\"odinger, \emph{Discussion of Probability Relations
    between
    Separated Systems}, Proc. Cambridge Philos. Soc. \textbf{31}, 555 (1935)
\bibitem{schr2}  E. Schr\"odinger, \emph{Probability relations between separated
    systems}, Proc. Cambridge Philos. Soc. \textbf{32}, 446 (1936)
\bibitem{epr} A. Einstein, B. Podolsky, N. Rosen, \emph{Can Quantum-Mechanical
    Description of Physical Reality Be Considered Complete?}, Phys. Rev.
    \textbf{47}, 777 (1935)
\bibitem{wis} S. J. Jones, H. M. Wiseman, A. C. Doherty, \emph{Entanglement,
    Einstein-Podolsky-Rosen correlations, Bell nonlocality, and steering}, Phys.
    Rev. A \textbf{76}, 052116 (2007)
\bibitem{piani} M. Piani, J. Watrous, \emph{Necessary and Sufficient Quantum
    Information Characterization of Einstein-Podolsky-Rosen Steering}, Phys. Rev.
    Lett. \textbf{114}, 060404 (2015)
\bibitem{branc} C. Branciard, E. G. Cavalcanti, S. P. Walborn, V. Scarani, H.
    Wiseman, \emph{One-sided device-independent quantum key distribution:
    Security,
    feasibility, and the connection with steering}, Phys. Rev. A \textbf{85},
    010301(R) (2012)
\bibitem{ser} A. Serafini, \emph{Quantum continuous variables: A primer of
    theoretical methods}, Taylor \& Francis Group (2017)
\bibitem{kog} I. Kogias, A. R. Lee, S. Ragy, G. Adesso, \emph{Quantification of
    Gaussian Quantum Steering}, Phys. Rev. Lett. \textbf{114}, 060403 (2015)
\bibitem{oliv} S. Olivares, \emph{Quantum optics in the phase space}, Eur. Phys.
    J.
    Spec. Top. \textbf{203}, 3 (2012)
\bibitem{ferr} A. Ferraro, S. Olivares, M. G. A. Paris, \emph{Gaussian states in
    continuous variable quantum information}, Bibliopolis, Napoli (2005)
\bibitem{weed} C. Weedbrook, S. Pirandola, R. Garcia-Parton, N. L. Cerf, T. C.
    Ralph, J. H. Shapiro, S. Lloyd, \emph{Gaussian quantum information}, Rev.
    Mod.
    Phys. \textbf{84}, 621 (2012)
\bibitem{ji} S.-W. Ji, J. Lee, J. Park, H. Nha, \emph{Steering criteria via
    covariance matrices of local observables in arbitrary-dimensional quantum
    systems}, Phys. Rev. A \textbf{92}, 062130 (2015)
\bibitem{frig} M. Frigerio, C. Destri, S. Olivares, M. G. A. Paris, \emph{Quantum
    steering with Gaussian states: A tutorial}, Phys. Lett. A \textbf{430},
    127954
    (2022)
\bibitem{hyll} P. Hyllus, J. Eisert, \emph{Optimal entanglement witnesses for
    continuous-variable system}, New J. Phys. \textbf{8}, 51 (2006)
\bibitem{janet} J. Anders, \emph{Estimating the degree of entanglement of unknown
    Gaussian states}, Diploma Thesis, University of Potsdam (2003),
    arxiv:quant-ph/0610263
\bibitem{tamih} T. Mihaescu, H. Kampermann, G. Gianfelici, A. Isar, D. Bruss,
    \emph{Detecting entanglement of unknown continuous
variable states with random measurements}, New J. Phys. \textbf{22}, 123041
(2020)
\bibitem{ma} R. Ma, T. Yan, D. Wu, X. Qi, \emph{Steering witnesses and steering
    criterion of Gaussian states}, Entropy \textbf{24}, 62 (2022)
\bibitem{sim} R. Simon, N. Mukunda, B. Dutta, \emph{Quantum-noise matrix for
    multimode systems: $U(n)$ invariance, squeezing, and normal forms}, Phys. Rev
    A
    \textbf{49}, 1567 (1994)
\bibitem{fur} J. Fiur\'a\ifmmode \check{s}\else \v{s}\fi{}ek, \emph{Gaussian
    Transformations and Distillation of Entangled Gaussian States}, Phys. Rev.
    Lett.
    \textbf{89}, 137904 (2002)
\bibitem{zhan} F. Zhang, \emph{The Schur complement and its applications},
    Springer
    (2005)
\bibitem{lami}  L. Lami, A. Serafini, G. Adesso, \emph{Gaussian entanglement
    revisited}, New J. Phys. \textbf{20}, 023030 (2018)
\bibitem{dut} A. B. Dutta, N. Mukunda, R. Simon, \emph{The real symplectic groups
    in
    quantum mechanics and optics}, Pramana-J. Phys. \textbf{45}, 471 (1995)
\bibitem{wolf} R. F. Werner, M. M. Wolf, \emph{Bound entangled Gaussian states},
    Phys. Rev. Lett. \textbf{86}, 3658 (2001)
\bibitem{will} J. Williamson, \emph{On the algebraic problem concerning the
    normal
    forms of linear dynamical systems}, Amer. J. Math. \textbf{58}, 141 (1936)
\bibitem{bathia} R. Bhatia, T. Jain, \emph{On symplectic eigenvalues of positive
    definite matrices}, J. Math. Phys. \textbf{56}, 112201 (2015)
\bibitem{dauria}V. D'Auria, A. Porzio, S. Solimeno, S. Olivares, M. G. A. Paris,
    \emph{Characterization of bipartite states using a single homodyne detector},
    J.
    Opt. B.: Quantum Semiclass. Opt. \textbf{7}, S750 (2005)
\bibitem{plen} Plenio, M. B.,
    \emph{Logarithmic Negativity: A Full Entanglement Monotone That is not
    Convex}, Phys. Rev. Lett. \textbf{95}, 119902 (2005)
\bibitem{jagger} D. P. Jagger, MATLAB toolbox for classical matrix groups, M. Sc.
    Thesis, University of Manchester (2003)
\bibitem{lock} P.  van Loock, S. L. Braunstein, \emph{Multipartite Entanglement
    for Continuous Variables: A Quantum Teleportation Network}, Phys. Rev. Lett.
    \textbf{84}, 3482 (2000)
\bibitem{addss} G. Adesso, A. Serafini, F. Illuminati, \emph{Three-mode Gaussian
    states in quantum information with continuous variables},  	
https://doi.org/10.48550/arXiv.quant-ph/0609071 (2006)
\bibitem{lock2} P. van Loock, S. L. Braunstein, \emph{Greenberger-Horne-Zeilinger
    nonlocality in phase space}, Phys. Rev. A \textbf{63}, 022106 (2001)
\bibitem{chi} K. Knight, \emph{Mathematical Statistics}, Chapman and Hall, New
    York
    (2000)
\bibitem{mor} T. Moroder, M. Kleinmann, P. Schindler, T. Monz, O. G\"uhne, R.
    Blatt,
    \emph{Certifying experimental errors in quantum experiments}, Phys. Rev.
    Lett.
    \textbf{110}, 180401 (2013)
\bibitem{toth} O. G\"uhne, G. Toth, \emph{Entanglement detection}, Phys. Rep.
    \textbf{474}, 1 (2009)


\end{thebibliography}
\end{document}